\documentclass{article}
\usepackage[T1]{fontenc}
\usepackage{wrapfig}
\usepackage[portuges,english]{babel}
\usepackage{amssymb,latexsym,amsmath,color,mathrsfs,ifsym,graphics,stmaryrd} 
\usepackage[colorlinks,linkcolor=blue,urlcolor=blue,citecolor=black,
plainpages=false,pdfpagelabels,breaklinks]{hyperref}

\title{Quantum Superpositions and the \\Representation of Physical Reality\\Beyond Measurement Outcomes\\and Mathematical Structures}
\author{{\sc Christian de Ronde}\thanks{Fellow Researcher of the Consejo
Nacional de Investigaciones Cient\'{\i}ficas y T\'ecnicas and Adjoint Professor of the National University Arturo Jaurteche.}}
\date{\begin{center}
\begin{small} 
CONICET, Buenos Aires University - Argentina \\
Center Leo Apostel and Foundations of  the Exact Sciences\\
Brussels Free University - Belgium \\
\end{small}
\end{center}}

\usepackage[margin=2.5cm]{geometry}

\begin{document}
\maketitle

\begin{abstract}
\noindent In this paper we intend to discuss the importance of providing a physical representation of quantum superpositions which goes beyond the mere reference to mathematical structures and measurement outcomes. This proposal goes in the opposite direction to the project present in orthodox contemporary philosophy of physics which attempts to ``bridge the gap'' between the quantum formalism and common sense ``classical reality'' ---precluding, right from the start, the possibility of interpreting quantum superpositions through non-classical notions. We will argue that in order to restate the problem of interpretation of quantum mechanics in truly ontological terms we require a radical revision of the problems and definitions addressed within the orthodox literature. On the one hand, we will discuss the need of providing a formal redefinition of superpositions which captures explicitly their contextual character. On the other hand, we  will attempt to replace the focus on the measurement problem, which concentrates on the justification of measurement outcomes from ``weird'' superposed states, and introduce the superposition problem which focuses instead on the conceptual representation of superpositions themselves. In this respect, after presenting three {\it necessary conditions} for objective physical representation, we will provide arguments which show why the classical (actualist) representation of physics faces severe difficulties to solve the superposition problem. Finally, we will also argue that, if we are willing to abandon the (metaphysical) presupposition according to which `Actuality = Reality', then there is plenty of room to construct a conceptual representation for quantum superpositions.
\medskip
\end{abstract}


\renewenvironment{enumerate}{\begin{list}{}{\rm \labelwidth 0mm
\leftmargin 0mm}} {\end{list}}

\newcommand{\ita}{\textit}
\newcommand{\mcal}{\mathcal}
\newcommand{\mfrak}{\mathfrak}
\newcommand{\mbb}{\mathbb}
\newcommand{\mrm}{\mathrm}
\newcommand{\msf}{\mathsf}
\newcommand{\mscr}{\mathscr}
\newcommand{\lra}{\leftrightarrow}
\renewenvironment{enumerate}{\begin{list}{}{\rm \labelwidth 0mm
\leftmargin 5mm}} {\end{list}}

\newtheorem{dfn}{\sc{Definition}}[section]
\newtheorem{thm}{\sc{Theorem}}[section]
\newtheorem{lem}{\sc{Lemma}}[section]
\newtheorem{cor}[thm]{\sc{Corollary}}
\newcommand{\Proof}{\textit{Proof:} \,}
\newcommand{\cqd}{{\rule{.70ex}{2ex}} \medskip}

\bigskip

\bigskip

\bigskip

\section*{Introduction}

Quantum superpositions are being used today in laboratories all around the world in order to create the most outstanding technological and experimental developments of the last centuries. Indeed, quantum computation, quantum teleportation, quantum cryptography and the like technologies are opening up an amazing range of possibilities for the near future. This new quantum technological era is founded on one of the main principles of Quantum Mechanics (QM), the so called {\it superposition principle} which in turn gives rise to {\it quantum superpositions} and {\it entanglement}. However, while many experimentalists are showing that Schr\"odinger's cats are growing fat \cite{Blatter00}, and it becomes increasingly clear that quantum superpositions are telling us something about quantum physical reality even at the macroscopic scale \cite{Nature15, NimmrichterHornberger13}, the philosophy of physics community does not seem interested in finding a coherent conceptual representation of them. Apart from very few non-mainstream proposals which will be discussed in the present article, there seems to be almost no attempt to create a conceptual explanation of the physical meaning of quantum superpositions which goes beyond the formal reference to a mathematical structure or the empirical reference to measurement outcomes.  

In this paper, we will argue that the reason for the lack of conceptual analysis regarding quantum superpositions is directly related to the exclusive emphasis that has been given in the orthodox literature to the infamous measurement problem. Taking as a standpoint a representational realist stance, we will argue for the need of replacing the measurement problem, which only focuses on the justification of observable measurement outcomes, by the superposition problem, which concentrates instead on the conceptual account of the mathematical expression itself. In order to discuss the possibility of interpreting superpositions we will present three {\it necessary conditions} which attempt to constrain any objective conceptual representation of an empirically adequate mathematical formalism. Our analysis will require a formal redefinition of the notion of quantum superposition which considers not only the meaningful physical statements that can be derived from them, but also their contextual character as basis-dependent formal elements of the theory. Taking into account our proposed three {\it necessary conditions} for the objective representation of physical reality, we will provide arguments which show the severe difficulties faced by the classical (actualist) representation of physics in order to produce a conceptual account of quantum superpositions. In the final part of the paper, we will also argue that, if we are willing to discuss the possibility that `Quantum Physical Reality $\neq$ Actuality', then there is plenty of room to represent quantum superpositions in terms of (non-actual) physical reality.

\section{The Representational Realist Stance}

Within philosophy of physics, realism has been characterized as a stance which assumes the existence of a reality independent of the actions of any human subject or conscious being. In short, it is claimed that realism is committed to the belief of an objective (subject independent) reality. However, we will argue that this account remains insufficient when attempting to grasp the {\it praxis} of realist physicists themselves. Representational realism attempts to capture exactly this aspect; i.e., the specific way through which realist physicists {\it produce} such representational (meta-physical) account of reality \cite{deRonde16b}. In this respect, the main presupposition of representational realism is that physical theories relate to reality, not only through their mathematical formalisms, but also through a net of physical concepts. Even though this characterization and understanding of physical theories is not new in the history of physics ---Einstein, Heisenberg, Pauli and Schr\"odinger, between many others, shared a similar perspective---, we believe it is important to restate realism beyond its present anti-metaphysical empiricist characterization ---inherited from the logico-positivist school of thought--- which grounds the development and understanding of theories in observable events ---considered as a ``common sense'' givens. Let us discuss this in more detail. 

The coherent interrelation between mathematical structures and physical concepts allows physical theories to represent (in various ways) the world and reality; it allows the physicist to imagine different physical phenomena. For example, in Newtonian mechanics the notions of space, time, inertia, particle, force, mass, etc., are related to infinitesimal calculus in such a way that we physicists are capable to imagine and predict very different phenomena which range from the motion of planets to the free fall of an apple from a tree. But infinitesimal calculus does not wear Newton's physical notions on its sleeves. There is no notion of `inertia', of `particle' or `absolute space' which can be read off the mathematical formalism of the theory. Mathematical equations have no physical concepts hidden within. `Space', `time', `particles', etc. are not ---according to our viewpoint---  self evident {\it givens} of experience; they are concepts that have been developed by physicists and philosophers through many centuries.\footnote{See for example, in this same respect, the detailed analysis of the concept of space in the history of physics provided by Max Jammer in his excellent book: {\it The Concepts of Space. The History of Theories of Space in Physics} \cite{Jammer93}.}

Every theory possesses not only its own mathematical formalism but also its own specific set of physical notions. As we all know, Maxwell's electromagnetic theory has a mathematical structure and set of concepts different to that of classical mechanics or Einstein's relativity theory. The fact that some theories share the same names for distinct physical concepts should not confuse us, their meaning in many cases differ radically. As remarked by Heisenberg \cite[pp. 97-98]{Heis58}: ``New phenomena that had been observed could only be understood by new concepts which were adapted to the new phenomena. [...] These new concepts again could be connected in a closed system. [...] This problem arose at once when the theory of special relativity had been discovered. The concepts of space and time belonged to both Newtonian mechanics and to the theory of relativity. But space and time in Newtonian mechanics were independent; in the theory of relativity they were connected.'' From this acknowledgment, the representational realist argues that the task of a realist physicist or philosopher of physics must be the creation of a closed theory which coherently interrelates a specific mathematical formalism, a particular net of physical concepts and a particular field of phenomena and experience described by that theory.

According to our representational realist stance, which will be presupposed through the rest of the paper, `experimental data' and `observability' should be regarded only in terms of a confirmation procedure about the empirical adequacy (or not) of a given theory. Our realist stance ---which relates to Heisenberg's closed theory approach\footnote{For a detailed discussion of the closed theory approach see \cite{Bokulich04, deRonde16b}.} and some of the main elements of Einstein's philosophy of physics\footnote{See: \cite{Howard93, Howard10}.}--- assumes that physical observation is both theoretically and metaphysically laden. In this respect, the basic principles of classical logic and metaphysics, namely, the Principle of Existence (PE), the Principle of Identity (PI) and the Principle of Non-Contradiction (PNC) determine, right from the start, the possibilities and limits of classical observation itself. Classical observation is both defined and constrained by these ontological and logical principles. Identities or non-contradictory properties are not something we find ``outside in the world'' ---no one has ever seen identity or non-contradiction walking in the street---, they are instead the basic conditions of possibility for discussing about any classical object of experience. Exactly the same point was made already by Hume himself regarding the physical notion of causality. As we all know, causation is not something grounded empirically, it is never found in the observable world. Rather, as Hume clearly exposed, it is a metaphysical presupposition which allows us to impose a specific sense to our experience.

PE, PNC and PI have been, not only the fundamental cornerstones of both Aristotle's metaphysics and classical logic itself, but also the main principles behind the physical picture imposed by Newton's classical mechanics (see \cite{VerelstCoecke}). The idea of ``classical reality'', which encompasses the whole of classical physics  (including relativity), can be condensed in the notion of {\it Actual State of Affairs} (ASA).\footnote{See \cite{RFD14} for discussion and definition of this notion in the context of classical physics.} This particular (metaphysical) representation was developed by Newton and can be formulated in terms of {\it systems constituted by a set of actual (definite valued) preexistent properties.} Thus, objects and their properties are defined through PE, PNC and PI as existents in the actual mode of being. In the 18th century these same principles constituted the basis for the metaphysical definition of the notion of actual entity in classical physics. Newton conceived a Universe constituted by bodies with always definite valued (actual) properties. This allowed Laplace to imagine a demon which, given the complete and exact knowledge of all particles in the Universe (i.e., the set of all systems and their actual properties), would have access, through the equations of motion, to the future and the past of the whole Universe. 

\begin{quotation}
\noindent {\small ``We may regard the present state of the universe as the effect of its past and the cause of its future. An intellect which at a certain moment would know all forces that set nature in motion, and all positions of all items of which nature is composed, if this intellect were also vast enough to submit these data to analysis, it would embrace in a single formula the movements of the greatest bodies of the universe and those of the tiniest atom; for such an intellect nothing would be uncertain and the future just like the past would be present before its eyes.'' \cite[p. 4]{Laplace}}\end{quotation}

Physical representation allows us to think about experience and predict phenomena without the need of performing any actual measurement. It allows us to imagine physical reality beyond the here and now. This is of course the opposite standpoint from empiricists who argue instead that the fundament of physics is `actual experimental data'. As remarked by Bas van Fraassen \cite[pp. 202-203]{VF80}: ``To develop an empiricist account of science is to depict it as involving a search for truth only about the empirical world, about what is actual and observable.'' In this case, ``actual'' has a very different meaning to the one just mentioned above. It is understood as making reference to {\it hic et nunc} self evident {\it actual observations} by empirical subjects, and not to the Aristotelian metaphysical definition of the {\it actual mode of existence} ---which is completely independent and transcends the existence of particluar empirical subjects. While the empiricist considers that the construction of theories always begins from observable data, the representational realist recognizes that theory construction is an entangled process of production of concepts, mathematical structures and fields of phenomena. Following Heisenberg \cite[p. 63]{Heis71}: ```Understanding' probably means nothing more than having whatever ideas and concepts are needed to recognize that a great many different phenomena are part of coherent whole. Our mind becomes less puzzled once we have recognized that a special, apparently confused situation is merely a special case of something wider, that as a result it can be formulated much more simply. The reduction of a colorful variety of phenomena to a general and simple principle, or, as the Greeks would have put it, the reduction of the many to the one, is precisely what we mean by `understanding'. The ability to predict is often the consequence of understanding, of having the right concepts, but is not identical with `understanding'.''  Contrary to the positivist-empiricist stance according to which observables in physics must be accepted as {\it givens} of experience, the representational realist considers observability as both theoretically and metaphysically laden. For her, the conceptual scheme is a {\it necessary condition} to produce not only the physical representation of a closed theory but also the condition of possibility to provide a categorical description of the (closed) physical phenomena discussed by that theory. The task of both physicists and philosophers is to jointly construct new mathematical formalisms and networks of physical concepts which allow us to imagine new physical phenomena. 

According to our stance, individual subjects (also called agents, users, etc.) should play no role in the description of physical reality. As Einstein \cite[p. 175]{Dieks88a} made the point: ``[...] it is the purpose of theoretical physics to achieve understanding of physical reality which exists independently of the observer, and for which the distinction between `direct observable' and `not directly observable' has no ontological significance'' even though ``the only decisive factor for the question whether or not to accept a particular physical theory is its empirical success.'' For the representational realist, empirical adequacy is part of a {\it verification procedure}, not that which ---according to van Fraassen \cite{VF80}--- needs to be saved. Following Einstein's dictum which allowed Heisenberg to derive his indeterminacy relations ``It is only the theory which decides what we can observe'' \cite[p. 63]{Heis71}. It is only the theory which provides the constraints that allows us to describe and understand what is observable. It is not, as argued by empiricists, the other way around.

This understanding of observability as being theoretically and metaphysically dependent goes not only against empiricism and instrumentalism but also against scientific realism itself. As Musgrave \cite[p. 1221]{PS} explains: ``In traditional discussions of scientific realism, common sense realism regarding tables and chairs (or the moon) is accepted as unproblematic by both sides. Attention is focused on the difficulties of scientific realism regarding `unobservables' like electrons.'' The main danger of all these philosophical positions ---namely, scientific realism, empiricism and instrumentalism--- is that they close the door to the development of new physical representations, since they assume that we already know what reality {\it is} in terms of the (naive) observation of tables and chairs ---also known as ``classical reality''. `Tables' and `chairs' are naively considered as ``common sense'' givens of experience and represented by mathematical models. Within these philosophical positions, contrary to representational realism, there is no reference whatsoever to metaphysical considerations regarding the specific meaning of `tables' and `chairs'. The orthodox project is then focused in trying ``to bridge the gap'' between the strange quantum formalism and ``classical reality'' \cite{Dorato15}. They attempt to do so, forgetting that our present ``common sense'' understanding of the world was also part of a creative scientific process, and not the final conditions of human understanding. Many years ago Einstein had stressed the importance of physical concepts and warned us of the dangerous threat of remaining captive of ``common sense realism'':  

\begin{quotation}
\noindent {\small ``Concepts that have proven useful in ordering things easily achieve such an authority over us that we forget their earthly origins and accept them as unalterable givens. Thus they come to be stamped as `necessities of thought,' `a priori givens,' etc. The path of scientific advance is often made impossible for a long time  through such errors. For that reason, it is by no means an idle game if we become practiced in analyzing the long common place concepts and exhibiting those circumstances upon which their justification and usefulness depend, how they have grown up, individually, out of the givens of experience. By this means, their all-too-great authority will be broken. They will be removed if they cannot be properly legitimated, corrected if their correlation with given things be far too superfluous, replaced by others if a new system can be established that we prefer for whatever reason.'' \cite[p. 102]{Einstein16}}
\end{quotation}

Empirically adequate mathematical structures are not enough to produce a meaningful physical representation of reality. Physics cannot be exclusively reduced to mathematical models which predict measurement outcomes simply because neither mathematical models nor empirical facts contain in themselves the physical concepts the theory talks about. The experience of looking to a `chair' or a `table' does not produce by itself the PI or the PNC. In fact, only by presupposing such principles we can say we observe a `chair' or a `table' as an individual constituted by non-contradictory properties. There are no physical concepts to be found within mathematical models either. Every physical theory is intrinsically constituted through specific physical concepts which are defined in categorical metaphysical terms ---through the systematization of specific principles. It is only through these representations that the experience described by a physical theory is made possible. As Heisenberg \cite[p. 264]{Heis73} makes the point: ``The history of physics is not only a sequence of experimental discoveries and observations, followed by their mathematical description; {\it it is also a history of concepts.}  For an understanding of the phenomena the first condition is the introduction of adequate concepts. {\it Only with the help of correct concepts can we really know what has been observed.}''\\

\noindent {\it {\bf Representational Realism:} A representational realist account of a physical theory must be capable of providing a physical representation of reality in terms of a network of physical concepts which coherently relates to the mathematical formalism of the theory and allows us to articulate and make predictions of definite phenomena. Observability in physics is always theoretically and metaphysically laden, and thus, it must be regarded as a consequence of each particular physical representation.}

\section{Representing Quantum Physical Reality}

For the representational realist, the task of both physicists and philosophers of physics is to produce conceptual representations which allow us to grasp, to imagine and understand the features of the Universe beyond mathematical schemes and measurement outcomes. In order to provide such representation we must necessarily complement mathematical formalisms with networks of physical concepts. It is simply not enough to claim that ``according to QM the structure of the world is like Hilbert space'', or that ``reality, according to QM is described through the quantum wave function''. That is just mixing the formal and conceptual levels of discourse (see for discussion \cite[Section 4]{RFD14}). That is not doing the job of providing a conceptual physical representation in the sense discussed above. Pure mathematics is simply incapable of producing physical concepts. To find adequate concepts and representations that explain what QM is talking about in not the task of mathematicians, it is the task of both physicists and philosophers of physics. 

If we accept the representational realist challenge and we are willing to discuss the conceptual meaning of QM, there seems to be two main lines of research to consider. The first one is to investigate the possibility that QM makes reference to the same physical representation provided by classical physics; i.e. that it talks about an ASA ---or in other words, about ``classical reality''. This is the main idea behind, for example, the hidden variables program which, as noticed by Guido Bacciagaluppi  \cite[p. 74]{Bacciagaluppi96}, attempts to ``restore a classical way of thinking about {\it what there is}.'' This attempt has been faced with severe difficulties, mainly due to the empirical test of Bell inequalities, the Kochen-Specker theorem, the infamous measurement problem, and the list continues. Due to the impossibility to account for a non-contextual classical representation, this line of research has abandoned the orthodox formalism and proposed instead new mathematical models which attempt to rescue ``classical reality'' from the weirdness of the quanta. The second line of research stays close to the orthodox formalism and considers the possibility that QM might describe physical reality in a different ---maybe even incommensurable--- way to that of classical physics. It is this line of research that interests us in this paper. 

In order to discuss and analyze physical interpretations we need to agree on the definition of what should be considered a Meaningful Operational Statements (MOS) within a theory. 

\begin{dfn}
{\sc Meaningful Operational Statements:} Every operational statement within a theory capable of predicting the outcomes of possible measurements must be considered as meaningful with respect to the representation of physical reality provided by that theory.
\end{dfn}
 
\noindent From a realist perspective MOS must be related, in the final stage of a theory construction, to the representation of reality provided by the theory. The intuition behind this requirement is remarked by Robert Griffiths \cite[p. 361]{Griffiths02}:  ``If a theory makes a certain amount of sense and gives predictions which agree reasonably well with experimental or observational results, scientists are inclined to believe that its logical and mathematical structure reflects the structure of the real world in some way, even if philosophers will remain permanently skeptical.'' As we argued above, the representation must be established not only in formal mathematical terms but also in conceptual terms through the introduction of appropriate physical notions. MOS are instrumentalist predictive statements about specific physical situations of the type: {\it if we prepare the experimental situation X, we will observe the measurement outcome Y}. It is the task of both physicists and philosophers of physics to complement MOS with adequate physical notions, allowing us to construct a coherent conceptual and categorical representation which allows us to explain what is exactly going on within the addressed phenomenon. 

The problem introduced by QM is that, even though we possess MOS such as, for example, `the spin of the quantum particle will be found to be + with probability 0.5' or `the atom will be found to be decayed with probability 0.7', we do not possess an adequate set of concepts that would allow us to grasp the physical content of these statements in relation to the present tense ---not in the same way as we do in Newton's mechanics or Maxwell's theory of electromagnetism. Let us be clear about this point, in QM we do not understand conceptually what is the meaning of the notions commonly used within the discourse of physicists in order to refer to the quantum formalism. The notions of `quantum particle', `spin', `atom' or even `probability' lack a categorical definition. Still today, we cannot picture what is gong on according to QM. When asked about such notions we can only shift to a ``mathematical explanation'' or recall the predictive empirical success of the theory. But prediction is not explanation. And neither is mathematical formalization. Prediction and formalization do not allow us to imagine what these ``quantum notions'' refer to. The idea that it is not possible to provide a representation of quantum reality  and that we should remain content with an instrumentalist account of QM has become, since Bohr's deep influence in the physics community,\footnote{According to Bohr \cite[p. 7]{WZ}: ``[...] the unambiguous interpretation  of any measurement must be essentially framed in terms of classical physical theories, and we may say that in this sense the language of Newton and Maxwell will remain the language of physicists for all time.'' In this respect, he aso added [{\it Op. cit.}, p. 7] that, ``it would be a misconception to believe that the difficulties of the atomic theory may be evaded by eventually replacing the concepts of classical physics by new conceptual forms.''} a silent dogma within the field of quantum foundations.  

In order to come up with a conceptual representation which explains the MOS of a theory, the produced physical discourse must allow for Counterfactual Reasoning (CR). This is due to the fact that counterfactual discourse  is a necessary condition for the possibility of a subject independent representation of reality. CR is used and analyzed by different disciplines. In the case of logicians and philosophers of science, CR is studied in terms of Kripke semantics, or possible worlds semantics. Even though this logical approach to counterfactuals has become popular in philosophy of QM (e.g. \cite{Griffiths02}), it has never been popular among physicists in general. In fact, physicists have always used counterfactuals in a rather (undefined) intuitive way in order to discuss physical experience as related to an objective description of reality. Let us provide thus a general definition of counterfactual reasoning which attempts to consider the actual {\it praxis} of physicists themselves.

\begin{dfn}
{\sc Counterfactual Reasoning:} If the theory is empirically adequate then the MOS it provides must be related to physical reality through a conceptual scheme. Counterfactual discourse is a necessary requirement in order to produce an objective physical representation of reality which is able to consider all the MOS the theory predicts. Due to CR, MOS escape their reference to the future and can also become statements about past and present events. Thus CR about MOS comprises all actual and non-actual observations. CR is the objectivity condition for the possibility of physical discourse. 
\end{dfn}

CR is an indispensable element for the physical discourse of a theory which attempts to discuss an objective representation of physical reality. Many of the most important debates in the history of physics have been related to thought experiments which make always explicit use of counterfactual discourse. In the 18th century, Newton and Leibniz imagined different physical situations, impossible to actually experience, in order to draw conclusions about classical mechanics. At the beginning of the 20th century, Einstein's famous {\it Gedankenexperiments} in relativity theory made clear that the notion of {\it simultaneity} in Newtoninan mechanics had to be reconsidered, producing a revolution in our understanding of space and time. During the 1920s Solvay meetings, Bohr and Einstein discussed in depth many {\it Gedankenexperiments} related to QM. Some years later, Schr\"odinger imagined a strange experimental situation in which an atom would get entangled to a superposed (quantum) cat  `dead' and `alive' at the same time. More than 50 years had to pass by in order to empirically test the existence of quantum superpositions and entanglement allowing technicians and experimentalists to explore amazing possibilities for quantum information processing. Also Einstein, Podolsky, Rosen and Bell had to wait till the 80s for Aspect and his group in order to be certain that the hidden variable project ---with which they wanted to replace QM by a ``complete theory''--- was not going to work out without giving up either {\it reality} or {\it locality}. These few examples show the crucial role played by conceptual representation and CR within the {\it praxis} of physicists. 

If we assume a representational realist stance, the conceptual representation must be capable not only of conceptually explaining the MOS produced by the theory, it must also produce a discourse which respects CR. Without CR in physical discourse one cannot imagine objective reality nor experience beyond the here and now. CR allows us to state that ``if I performed this (or that) experiment'' then ---if it is, of course, a MOS--- the physical theory will tell me that ``the outcome will be $x$ (or $y$)'', and I do not need to actually perform the experiment! I know what the result will be independently of performing the experiment or not. And that is the whole point of being a realist about physics, that I trust the theory to be talking about a physical representation of reality of which I can make sense without making any reference to the {\it hic et nunc} observation of a particular subject. 

Physicists are accustomed to play with the counterfactual statements produced by a theory. CR in physical discourse has nothing to do with time, evolution nor dynamics, it has to do with the possibility of representing objective physical reality and experience. A physical theory allows me to make counterfactual statements about the future, the present or the past, just in the same way physical invariance in classical mechanics connects the multiple frames of reference without anyone actually being in any of them. CR is the discursive condition with respect to the objective physical description of phenomena. Indeed, in classical mechanics (or relativity theory), we do not need to actually {\it be} in a specific frame of reference to {\it know} what will happen in that specific frame, or a different one. We can imagine and calculate what will happen in each frame, we can physically represent them to ourselves and translate what will happen in each of them through the Galilean (or Lorentz) transformations. And this of course makes explicit use of CR. For a realist, the possibility to imagine and picture reality is the magic of physics. Once we believe to have an empirically adequate formalism, we realists ---contrary to empiricists--- still have a lot of work to do, we still need to produce a {\it representation} of what the theory is talking about before we can claim we understand observations. And exactly that, is what seems to be lacking in the case of QM.  

We physicists can imagine how a distant star will collapse and transform into a white dwarf many, many years from now; we can also understand what would happen to space and time when traveling on a ray of light; or the tremendous consequences of what could happen to us inside a back hole; we can even discuss what already happened 13.800 million years ago during the first minutes of the Universe after the Big Bang, long before any conscious being existed. It is the trust in the physical representations provided by different theories which allows physicists to imagine situations which escape not only the spacial region in which they live, but also the technical possibilities of their time. And that is the reason why, as Einstein remarked, imagination is more important than knowledge.
 
From the previous analysis, we propose the following three {\it necessary conditions} for producing an objective physical representation of a theory:

\begin{enumerate}
\item[{\bf Necessary Condition 1 ($NC_1$).}] {\it Every physical theory must be capable of producing MOS which can be empirically tested. }

\item[{\bf Necessary Condition 2 ($NC_2$).}] {\it Every MOS the theory produces must be directly related to the representation of physical reality, provided through a specific conceptual scheme which adequately explains the phenomenon.}

\item[{\bf Necessary Condition 3 ($NC_3$).}] {\it The conceptual representation of a physical theory must be capable of producing a coherent counterfactual physical discourse which includes all MOS of the theory.}
\end{enumerate}

\noindent In QM, the complementarity scheme produced by Bohr violates explicitly CR making impossible to provide an objective representation. As we have discussed elsewhere \cite{deRonde16b}, the orthodox Bohrian scheme faces severe difficulties in this respect. The demand to provide a realist conceptual representation of QM implies not only a different perspective, it also presupposes the consideration of new problems, one of which will be discussed explicitly in the following section.

\section{The Superposition Problem: Representation Beyond Actual \\Outcomes}

The orthodox line of research deals with a specific set of problems which analyze QM from a classical perspective. This means that all problems assume as a standpoint ``classical reality'' and only reflect about the formalism in ``negative terms''; that is, in terms of the failure of QM to account for the classical representation of reality and the use of its concepts: separability, space, time, locality, individuality, identity, actuality, etc. The ``negative'' problems are thus: {\it non-}separability, {\it non-}locality, {\it non-}individuality, {\it non-}identity, etc.\footnote{I am grateful to Bob Coecke for this linguistic insight. Cagliari, July 2014.} These ``no-problems'' begin their analysis considering the notions of classical physics, assuming implicitly as a standpoint the strong metaphysical presupposition that QM should be able to represent physical reality in terms of such classical notions. The most famous of all interpretational problems of QM, is the so called ``measurement problem''.\\

\noindent {\it {\bf Measurement Problem:} Given a specific basis (or context), QM describes mathematically a quantum state in terms of a superposition of, in general, multiple states. Since the evolution described by QM allows us to predict that the quantum system will get entangled with the apparatus and thus its pointer positions will also become a superposition,\footnote{Given a quantum system represented by a superposition of more than one term, $\sum c_i | \alpha_i \rangle$, when in contact with an apparatus ready to measure, $|R_0 \rangle$, QM predicts that system and apparatus will become ``entangled'' in such a way that the final `system + apparatus' will be described by  $\sum c_i | \alpha_i \rangle  |R_i \rangle$. Thus, as a consequence of the quantum evolution, the pointers have also become ---like the original quantum system--- a superposition of pointers $\sum c_i |R_i \rangle$. This is why the measurement problem can be stated as a problem only in the case the original quantum state is described by a superposition of more than one term.} the question is why do we observe a single outcome instead of a superposition of them?}\\

\noindent The measurement  problem is also a way of discussing the quantum formalism in ``negative terms''. In this case, the problem concentrates in the justification of observable measurement outcomes. It should be remarked that the measurement problem presupposes that the basis (or context) ---directly related to a measurement set up--- has been already determined (or fixed). Thus it should be clear that there is no question regarding the contextual character of the theory within this specific problem. As we have argued extensively in \cite{deRonde16b}, the measurement problem has nothing to do with contextuality. The problem raises when, within a definite basis, the actualization process is considered. There is then a mix of subjective and objective elements when the recording of the experiment takes place ---as Wigner and his friend clearly explained \cite{WZ}. The problem here regards the coherency between the physical representation provided when the measurement was not yet performed, and the system is mathematically described in terms of a quantum superposition; and when an experimentalist claims in the lab ``I have observed a single measurement outcome'', which is not described by the theory. Since there is no physical representation of ``the collapse'', the subject (or his friend) seems to define it explicitly. The mixture of objective and subjective is due to an incomplete description of the state of affairs within the quantum measurement process (or ``collapse''). 

The focus of the measurement problem is the {\it hic et nunc} actual realm of experience. In this sense, the measurement problem is an empiricist problem which presupposes the controversial idea that actual observation is perfectly well defined in QM. However, from a representational realist stance things must be analyzed in a radically different manner. Indeed, if we agree with Einstein that it is only the theory ---both formally and conceptually--- which can tell you what can be observed then the measurement outcome should be treated without assuming what might be considered as ``self evident'' or ``obvious'' presuppositions (e.g., because there is suddenly a spot appearing in a photographic plate this obviously means I observed an elementary particle). At this point our approach makes a radical move. If we are willing to truly investigate the physical representation of quantum superpositions, rather than trying to justify what we observe in ``common sense'' classical terms, we need to ``invert'' the measurement problem and focus on the formal-conceptual level. This means that the attention should be focused on the conceptual representation of the mathematical expression itself instead of attempting to somehow ``save'' the measurement outcomes justifying through {\it ad hoc} rules the ``collapse'' of the quantum superposition to one of its terms. 

The important developments we are witnessing today in quantum information processing demands us, philosophers of QM, to pay special attention to the novel requirements of this new technological era. In this respect, we believe that a task of outmost importance is to produce a coherent physical representation of quantum superpositions.\\

\noindent {\it {\bf Superposition Problem:} Given a situation in which there is a quantum superposition of more than one term, $\sum c_i \ | \alpha_i \rangle$, and given the fact that each one of the terms relates through the Born rule to a MOS, the question is how do we conceptually represent this mathematical expression? Which is the physical notion that relates to each one of the terms in a quantum superposition?}\\
 
\noindent The superposition problem opens the possibility to truly discuss a physical representation of reality which goes beyond the classical representation of physics in terms of a Newtonian ASA, or its mere reference to measurement outcomes and mathematical structures. It clarifies what needs to be done for a convinced representational realist. We believe that without a replacement of the measurement problem by the superposition problem, there is no true possibility of discussing an interpretation of QM which provides an objective non-classical physical representation of reality. We know of no reasons to believe that this is not doable.

\section{Contextual Formal Definition of Quantum Superpositions}

Both the measurement and the superposition problems imply the necessary requirement that the quantum superposition is formally defined in a contextual manner. The fact that a quantum state $\Psi$ can be mathematically represented in multiple bases must be explicitly considered within such definition. This obvious remark might be regarded as controversial due to the fact the contextual character of quantum superpositions has been completely overlooked within the orthodox literature. As we have discussed in detail in \cite{daCostadeRonde16}, the semantic interpretation used in order to interpret the syntactical level of the quantum formalism presupposes implicitly PE, PNC and PI. This ``common sense''  classical interpretation has been uncritically accepted without considering the required necessary coherency between the addressed semantical and syntactical levels of the theory.  

Another consequence of this classical-type semantical interpretation is the complete omission of the obvious distinction between two different levels of mathematical representation present within the orthodox formalism. Indeed, the use of notions of `system', `state' and `property' has camouflaged the important distinction between, on the one hand, an abstract vector, $\Psi$, and on the other, its different basis-dependent representations, $\sum c_i \ | \alpha_i \rangle$. Let us remark again that this distinction is already implicit within the measurement problem itself. The measurement problem requires the necessary specification of the particular basis in which the state is represented as a linear combination of terms. Thus, in order to make explicit these two different formal levels, we state now the following contextual definition of quantum superpositions:

\begin{dfn}
{\sc Quantum Superposition:} Given a quantum state, $\Psi$, each i-basis defines a particular mathematical representation of $\Psi$, $\sum c_i \ | \alpha_i \rangle$, which we call a quantum superposition. Each one of these different basis dependent quantum superpositions defines a specific set of MOS. These MOS are related to each one of the terms of the particular quantum superposition through the Born rule. The notion of quantum superposition is contextual for it is always defined in terms of a particular experimental context (or basis).
\end{dfn}

The contextual character of quantum superpositions is an aspect of outmost importance when attempting to conceptually represent them. Let us discuss an explicit example in order to clarify these ideas. Consider a typical Stern-Gerlach type experiment where we have produced a quantum state $\Phi$ mathematically represented in the $x$-basis by the ket $| \uparrow_x>$. This can be easily done by filtering off the states $| \downarrow_x>$ of a Stern-Gerlach arranged in the $x$-direction. It is a mathematical fact that this state can be represented in different bases which diagonalyze a complete set of commuting observables. Each basis-dependent representation of the sate $\Phi$ is obviously different when considering its physical content. Indeed, it is common to the {\it praxis} of physicists to relate each different basis with a specific measurement context. 

The notion of basis, and thus also of superposition as we defined it above, possesses a physical content which relates a specific set of epistemic inquiries regarding the abstract state $\Phi$ to a set of MOS which provide an  answer to each specific question. All MOS are context-dependent. In our example, we know that if we measure  $\Phi$ in the context given by the $x$-basis we will observe with certainty `spin up'. Thus, the MOS related to the $| \uparrow_x>$ is of course: ``if the SG is in the $x$-direction then the result will be `+' with certainty (probability = 1)''. Physicists are taught that if they want to learn what are the possible outcomes in a different context, for example if they turn the SG to the $y$-direction, then they just need to write the state $| \uparrow_x>$ in the $y$-basis. Through this change of basis, and according to our previous definition, physicists obtain a different quantum superposition: $ c_{y1} | \uparrow_y> + \ c_{y2} | \downarrow_y>$. Writing the state in the $y$-basis produces a new superposition which relates to the following two MOS. The first one is that ``if the SG is in the $y$-direction then the result will be `+' with probability $|c_{y1}|^2$''. The second MOS is: ``if the SG is in the $y$-direction then the result will be `-' with probability $| c_{y2}|^2$''. The same will happen with any $i$-context of inquiry (determined by a particular $i$-basis), each one of them will be related to a particular quantum superposition $ c_{i1} | \uparrow_i> + \ c_{i2} | \downarrow_i>$ and a specific set of MOS arising from it. Following $NC_1$, all these different context-dependent MOS can be tested and provide empirical content to each one of the different superpositions. Thus, according to our $NC_2$ and $NC_3$, each one of these particular MOS must be related to physical reality through a conceptual physical representation which respects counterfactual discourse and reasoning. Following our previous analysis we will show that if we assume that there is only one world ---and not many, as Oxford Everettians believe---, quantum superpositions cannot be represented in terms of the actual mode of existence.

\section{Quantum Superpositions and the Actual Realm}

According to the three {\it necessary conditions} imposed by representational realism, each MOS, like the ones described in our Stern-Gerlach example of the previous section, must be related to a representation of physical reality. However, if we attempt to do so within the realm of actuality the conflict becomes evident. On the one hand, the realm of actuality ---determined through PE, PNC and PI--- cannot allow a physical description in terms of {\it contradictory properties}. On the other hand, QM produces, at least within some specific contexts, contradictory statements such as, for example, `the atom has the property of being decayed' and `the atom has the property of being not-decayed'. But obviously, within the actual realm of existence, an atom cannot posses both properties simultaneously, for that would flagrantly violate PNC. The concept of atom is defined as a substance which possesses non-contradictory properties. Thus, the atom must be necessarily {\it either} `decayed' or `not-decayed'.\footnote{See for a detailed discussion of the paraconsistent character of quantum superpositions: \cite{daCostadeRonde13}. Also, for a debate regarding the paraconsistent aspect of superpositions: \cite{ArenhartKrause15a, ArenhartKrause15b, deRonde15}.} 

The simplest way to escape this paradox would be to argue that the probabilities that accompany the states are {\it epistemic probabilities}. That would allow us to argue that the atom {\it is} in fact, {\it either} `decayed' or `not-decayed'; the problem would then be that we do not {\it know} which is the actual state of the atom ---implying that the atom has in fact a non-contradictory actual state. Unfortunatelly, orthodox QM precludes such epistemic interpretation of quantum probability (see e.g. \cite{Redei12}). As it is well known, the quantum probability arising from the orthodox formal structure via Gleason's theorem is non-Kolmogorovian and does not accept an epistemic interpretation. Quantum probability simply cannot be understood as making reference to our ignorance about an ASA. Also, from a more physical perspective the ``interaction'' of these possibilities makes it difficult not to consider all terms as relating to physical reality.  

Regardless of the formal constraints, quantum Bayesianism (QBism for short)  \cite{QBism13} does interpret quantum probability in epistemic terms, but this is done at the very high cost ---at least for a realist--- of having to claim that ``QM does not talk about physical reality'', that it is only ``an algorithm'' which only accounts for measurement observations of subjects, agents or users \cite[p. 71]{FuchsPeres00}. Following QBism, D'Ariano and Perinotti have explained in a recent paper \cite[pp. 280-281]{DArianoPerinotti16} that: ``What happens in the Schr\"odinger cat thought experiment is that the nonlocal test has no intuitive physical interpretation, since it is incompatible with all possible local observations. [...] But if one reasons operationally, it is evident that there is no logical paradox, and the described experiment is only highly counterintuitive.'' In fact, when reasoning in purely operational terms, the Schr\"odinger cat paradox cannot be posed. If one does not accept the premise that QM should be related to a representation of physical reality, then the realist question regarding the physical meaning of quantum superpositions ---exposed by the superposition problem--- cannot be addressed. This is not solving the problem, it is rather assuming a philosophical standpoint which invalidates the question regarding the conceptual meaning of quantum superpositions. 

An analogous interpretation would be to argue that quantum probabilities need to be interpreted in terms of future events (see e.g. \cite{Sudbery16}) or conscious observations (see e.g. \cite{Gao15}). These responses escape the question at stake for the problem of interpretation of quantum superpositions is not that of epistemic prediction, it is that of ontological representation. We are not discussing here whether QM predicts the correct measurement outcomes in an experiment ---we already know it does---, we want to understand how this occurs in terms of a conceptual physical representation. And this is why the physical explanation we seek requires necessarily a conceptual level. 

One of the first to see the difficulties of interpreting quantum superpositions was Paul Dirac who wrote in his famous book \cite{Dirac74}: ``The nature of the relationships which the superposition principle requires to exist between the states of any system is of a kind that cannot be explained in terms of familiar physical concepts. One cannot in the classical sense picture a system being partly in each of two states and see the equivalence of this to the system being completely in some other state.'' Also Schr\"odinger made very clear, in his famous 1935 paper \cite[p.  153]{Schr35}, the deep difficulties one is immersed in when attempting to represent quantum superpositions in terms of our classical (actualist) representation of physical reality: ``The classical concept of {\it state} becomes lost, in that at most a well-chosen {\it half} of a complete set of variables can be assigned definite numerical values.'' He also remarked that the problem cannot be solved by making reference to epistemic uncertainty or the measurement process. ``One should note that there was no question of any time-dependent changes. It would be of no help to permit the [quantum mechanical] model to vary quite `unclassically,' perhaps to `jump.' Already for the single instant things go wrong.'' As he made the point: ``[...] if I wish to ascribe to the model at each moment a definite (merely not exactly known to me) state, or (which is the same) to {\it all} determining parts definite (merely not exactly known to me) numerical values, then there is no supposition as to these numerical values {\it to be imagined} that would not conflict with some portion of quantum theoretical assertions.''\footnote{[{\it Op. cit.}, p. 156].}

The necessity of considering the multiple terms of a quantum superposition as physically real is supplemented by the fact that the terms `evolve' and `interact' according to the Schr\"odinger equation producing specific predictions which can be empirically tested and are in accordance to such `evolution' and `interaction'. As remarked by Everett \cite[p. 150]{Everett73} himself: ``It is [...] improper to attribute any less validity or `reality' to any element of a superposition than any other element, due to [the] ever present possibility of obtaining interference effects between the elements, all elements of the superposition must be regarded as simultaneously existing.'' According to our representational realist stance, such empirical findings must be necessarily related to a physical representation of reality. Some modal interpretations have attempted to escape this problem by arguing that one can relate these MOS to {\it possibilities} rather than {\it actualities} \cite{Bub97, Dieks07}. But the problem here is that in classical physics, possibilities never interact. Only actualities are allowed to evolve and interact with other actualities. That {\it possibilities can interact} is an idea that simply makes no sense within the classical actualist Newtonian representation of physical reality. In classical physics, possibilities are always epistemic possibilities. As Dieks clearly explains:

\begin{quotation}
\noindent {\small ``In classical physics the most fundamental description of a physical system (a point in phase space) reflects only the actual, and nothing that is merely possible. It is true that sometimes states involving probabilities occur in classical physics: think of the probability distributions $\rho$ in statistical mechanics. But the occurrence of possibilities in such cases merely reflects {\it our ignorance} about what is actual. The statistical states do not correspond to features of the actual system (unlike the case of the quantum mechanical superpositions), but quantify our lack of knowledge of those actual features. This relates to the essential point of difference between quantum mechanics and classical mechanics that we have already noted: in quantum mechanics the possibilities contained in the superposition state may interfere with each other. There is nothing comparable in classical physics. In statistical mechanics the possibilities contained in $\rho$ evolve separately from each other and do not have any mutual influence. Only one of these possibilities corresponds to the actual situation. The above (putative) argument for the reality of modalities can therefore not be repeated for the case of classical physics.'' \cite[pp. 124-125]{Dieks10}}\end{quotation}

\noindent The fact that quantum possibilities do interact is, according to us, the main mystery introduced by the theory of quanta. The Humean interpretation  proposed by Dieks in the same paper [{\it Op. cit.}] ---which only considers the future events predicted by QM--- seems to us an empiricist escape which does not address the representational realist superposition problem defined above. 

There are many interpretations which seem to shift the debate from the conceptual meaning of a mathematical element of the theory to the possibility of prediction given by that same formal element. But as we argued above, for a realist, prediction is not explanation. In fact, there exists a vast literature composed by two main lines of research which investigate the possibility to consider quantum superpositions from realist perspectives. The first line is related to the idea that each term in a superposition relates to the existence of a ``world'', ``branch'', ``mind'' or ``history''. Examples of these interpretations are Everett's many worlds interpretation \cite{MW?, Wallace12} the many minds variant proposed by Albert and Lower \cite{AlbertLower}, the consistent histories interpretation proposed by Griffiths \cite{Griffiths02} and the decoherent variant by Gell-Mann, Hartle \cite{Hartle15} and Omn\`es \cite{Omnes94}. Even though this family of interpretations might seem to provide an answer to the measurement problem arguing that actual reality is multiple, they still have two serious problems to confront. The first is the basis problem, which attempts to justify the subjective choice of a particular ``preferred'' basis between the many incompatible ones. The proposed solution to this problem in terms of the process of decoherence has found serious criticisms \cite{DawinThebault15, Kastner14}. The second problem is the interpretation of probability, which according to the orthodox formalism is incompatible with an epistemic interpretation. Following Everett's epistemic viewpoint regarding measurement, Deutsch and Wallace \cite{Deutsch99, Wallace07} have proposed to join QBism and interpret quantum probability as a subjective epistemic belief of ``rational agents'' ---or ``users'', as Mermin prefers to call them \cite{Mermin15}. This proposal also confronts very serious difficulties \cite{Jansson16}. In particular, if superpositions describe the co-existence of many different worlds then the quantum probability derived from the quantum superpositions themselves\footnote{The Born rule provides the probability of finding a certain observable via the numbers that accompany the kets within quantum superpositions.} should be directly related to the (ontological) existence of such worlds and not to the (epistemic) rational choice of individual agents described by Bayesian probability. Others like Griffiths, understand the probability related to the terms in a superposition simply as a ``tool'' to calculate outcomes \cite{Griffiths13}. The main problem surrounding the epistemic and instrumentalist interpretations of probability within supposedly realist interpretations of QM is that they provide no conceptual understanding of the weird interaction of probable states described by the theory. Once again, using a formal scheme that ``works'' and provides the correct measurement outcomes in probabilistic terms, is clearly very different from understanding and representing what is really going on according to the theory. The ontological question about {\it what there is} (independently of subjects) according to a theory obviously cannot be solved from an epistemic viewpoint which assumes that theories only make reference to the {\it prediction of observations} by individual subjects (agents or users). 

The second realist line of research investigates the idea of the existence of {\it indefinite properties} described in terms of propensities, dispositions, potentialities, possibilities or latencies. There are many different examples of this large family of interpretations. Let us mention at least some of them. Heisenberg's potentiality interpretation developed in terms of operational quantum logic by Piron \cite{Piron83},  Popper's propensity interpretation and Margenau's latency interpretation presently developed by Suarez, Dorato and Esfeld \cite{Dorato15}, the modal interpretations of Dieks and Bub \cite{Dieks88a, Bub97}. However, all these interpretations share a common difficulty. As remarked by Dorato himself with respect to dispositions: 

\begin{quotation}
\noindent {\small ``[...] dispositions express, directly or indirectly, those regularities of the world around us that enable us to predict the future. Such a predictive function of dispositions should be attentively kept in mind when we will discuss the `dispositional nature' of microsystems before measurement, in particular when their states is not an eigenstate of the relevant observable. In a word, the use of the language of `dispositions' does not by itself point to a clear ontology underlying the observable phenomena, but, especially when the disposition is irreducible, refers to the predictive regularity that phenomena manifest. {\it Consequently, attributing physical systems irreducible dispositions, even if one were realist about them, may just result in more or less covert instrumentalism.}'' \cite[p. 4]{Dorato06} (emphasis added)}\end{quotation}

\noindent This deep criticism to dispositions can be easily extended to the description of {\it indefinite properties} in terms of propensities, possibilities and potentialities. The reason is that the just mentioned interpretations end up defining propensities, possibilities and potentialities exactly in the same way as it is done by dispositionalists, namely, in terms of the future actualization of measurement outcomes. And for this reason, they all fall pray of Dorato's criticism. The lack of a (conceptual) categorical definition of these notions does not allow to imagine what these strange {\it indefinite properties} amount to beyond their predictive capacity regarding future observations. Many of these interpretations try to fill the representational lacuna using a metaphoric discourse which makes use of undefined meaningless notions such as ``elementary particle'', ``wave'', ``quantum filed'' or ``atom''. However, due to their incompatibility with the formalism, such notions are incapable of producing a coherent picture of what is really going on. Without conceptual representation there is no clear explanation of what physical reality amounts to according to the theory, and thus the main question remains also unanswered. 

Both of these general lines of research have concentrated their efforts in trying to solve the measurement problem. None of these general schemes break with the actualist understanding of physical reality. While the former line of research attempts to restore classicality by multiplying our world into a multiverse of many (un-observable) branching worlds, minds or histories; the latter introduces potential, propensity or dispositional type properties explicitly defined in terms of their future actualization. As remarked by Dorato this solution seems to result in nothing else than ``more or less covert instrumentalism''. Both lines have deep problems in order to meet the requirements of a representational realist project according to which the understanding of QM requires an explicit conceptual and categorical description of {\it what physical reality is} according to the theory beyond the mere reference to measurement outcomes or abstract mathematical structures. 

A detailed analysis of these interpretations exceeds the space of this article which we leave for a future work. In the present paper we attempt to consider a radically different path. That is, to address the question of how to extend the notion of reality in order to produce an objective description of physical reality in accordance with the orthodox formalism of QM. The price we are willing to pay is the abandonment of a metaphysical equation which has become a silent dogma within philosophy of physics, the idea that `Reality = Actuality'.

\section{Representing Quantum Superpositions Beyond the Actual Realm}

Today, experimentalists in laboratories are playing with superpositions and entanglement all around the world. Physicists are developing a new era of quantum technology far away from the measurement problem. It is in fact this {\it praxis} of physicists which should call our attention as philosophers of QM. How are quantum superpositions being treated by physicists in the lab? We believe this is an important question we should definitely consider. After more than one century of not being able to interpret QM in terms of the actualist representation of ``classical reality'', it might be time we admit that QM confronts us with the fact that the classical representation of physics might  not be the end of the road. We might be in need of abandoning physical representation exclusively in terms of ``classical reality'' (see for discussion \cite{deRonde16a, deRonde16b}). 

Quantum superpositions `evolve' and `interact' according to the Schr\"odinger equation of motion. Their MOS can be empirically tested in the lab through the Born rule ($NC_1$). But when in physics a mathematical element of a theory `evolves', `interacts' and `can be predicted' according to a mathematical formalism, then ---always from a representational realist perspective--- the elements of such mathematical expression need to be related (in some way) to specific physical notions which are capable of producing a coherent counterfactual discourse and representation of physical reality ($NC_2$ and $NC_3$). Newton was capable of explaining the movement of the planets in the sky and the bodies on Earth through the creation of specific concepts such as `inertia', `absolute space', `absolute time', `mass', etc. After centuries, these concepts became ---to great extent--- part of our ``common sense'' representation of the world. Also Maxwell was capable of explaining many different experiments and equations through the introduction of the notions of `electromagnetic filed' and `charge'. Physical theories are capable of producing not only a quantitative mathematical account of phenomena, they are also capable of producing a qualitative conceptual representation. In the case of QM, we know that each quantum superposition is related to a specific set of MOS within a particular measurement context. However, we do not understand what they {\it represent} in conceptual terms. We do not know what do they {\it refer to}. Superpositions impose a difficult dilema when attempting to interpret the orthodox formalism of QM. So it seems, either the formalism should be changed in order to restore ``classical reality'', or we should create a new understanding of physical reality itself beyond the constraints imposed by the classical actualist representation of physics ---which boil down to the actualist metaphysical scheme imposed by PE, PNC and PI. The latter path implies taking seriously the logical possibility that `Quantum Physical Reality $\neq$ Actuality'. 

Our representational realist stance seems to force us, given the predictions provided by QM, to extend the realm of what is considered to be real. Since both {\it certain} (probability equal to unity) and {\it statistical} (probability between zero and unity) predictions about physical quantities provide empirical knowledge, we believe there is no reason ---apart from dogmatism regarding actualist metaphysics--- not to relate both predictions to physical reality. This means we need to be creative enough to produce a new understanding of probability in terms of objective knowledge, abandoning its classical understanding in terms of ignorance about an ASA. If we accept the challenge of representational realism and admit that quantum superpositions must be related to a conceptual level of description, then there are two main mathematical elements we need to conceptually represent in terms of objective physical concepts. Firstly, we need to provide a clear representation of the kets that constitute each quantum superposition ---orthodoxly interpreted through their one-to-one relation to projection operators as properties of a quantum system. Secondly, we need to explain the physical meaning of the numbers that accompany the kets ---orthodoxly interpreted as related to the probability of finding the respective property. If we were able to extend the limits of what can be considered as physically real, we might be also able to open the door to a new understanding of QM beyond the orthodox classical reference to `systems', `states' and `properties'. 

There exist in the literature two interesting interpretations that might be considered to be advancing in the direction proposed in this article. The first is the conceptuality interpretation of QM, due to Diederik Aerts \cite{Aerts09, Aerts10}. The second is the development of the transactional interpretation due to Ruth Kastner \cite{Kastner12, Kastner15}. Both interpretations go in the direction of considering the main features of QM as central characteristics of the theory, rather than problems that need to bypassed. In the following we will discuss in what sense these interpretations might be able to answer the questions we have posed above. We will then continue to consider some elements of our own proposal.

Aerts interpretation \cite[p. 361]{Aerts09} is based on ``the hypothesis that a quantum particle is a conceptual entity, more specifically, that a quantum particle interacts with ordinary matter in a similar way than a human concept interacts with a memory structure.'' Aerts goes against the orthodox viewpoint according to which QM describes either particles or waves and attempts to show that ``if quantum entities are considered to be concepts rather than objects, and hence neither particles nor waves, the type of structure provoking entanglement and non-locality appears in a natural way.'' Through the introduction of the notion of `state of a concept' Aerts is able to show how linguistic concepts that are combined give rise to superpositions, entanglement, interference and non-locality. Contrary to the widespread animosity against quantum superpositions present in the orthodox foundational literature, Aerts attempts to provide an original explanation of what is really going on with superposed states. The main idea is that \cite[p. 27]{Aerts10} ``a quantum particle in a superposition state `is not inside space', and its `being inside space becomes only actualized due to a position measurement'. [...] If quantum particles are not inside space, it means that space is an emergent structure, coming into being jointly with the macroscopic material objects populating it and interacting in it.'' The example provided by Aerts in order to explain what is really going on in QM makes use of one of the most important type of interactions between human beings, namely, the communication through language.  

\begin{quotation}
\noindent {\small ``We believe that it is worth to explore the nature of the `interaction of human minds communicating with each other through language' so as to try to gain a better understanding of quantum mechanics and how it models the behavior of quantum particles interacting with matter. The reason is that we have recently been able to use the quantum mechanical formalism to model mathematically the interaction between human minds through language, which has strongly stimulated the investigation related to this new interpretation for quantum mechanics (Aerts, 2009; 2010).'' \cite[p. 31]{Aerts10}} \end{quotation}

\noindent There are two main stages within the quantum research directed by Aerts. The first stage is related to the quantum mechanical modeling of decision process. Taking advantage of the well known contextual character of the quantum formalism Aerts and Aerts could model in a satisfactory manner the way in which contexts in specific decision-making situations have an essential impact on the manner in which the decision is ultimately reached \cite{AertsAerts94}. The second stage consisted in ``showing that the quantum effects of `interference' and `superposition' could model very well so far unexplained and little understood but experimentally well-documented effects'' in cognition and decision theory such as the disjunction fallacy and the disjunction effect. Aerts and collaborators worked out a quantum-modeling scheme for the application of interference and superpositions for the aformentioned type of problems, fallacies and effects \cite{Aerts09b, Aerts09c, AertsHooghe09}. In particular, it is interesting to notice ``that `superposition' in this quantum-modeling described `the emergence of a new conceptual entity'. In the case of concepts and their combinations, this new conceptual entity is a new concept.'' \cite{Aerts09b, AertsHooghe09} According to Aerts:

\begin{quotation}
\noindent {\small ``[...] common objects do not appear in quantum superposition states, because there is no interface capable of interacting with them. [...] the superposition state of, for example, two chairs in different locations A and B, exists without being a state of a concept which can be interpreted as an object. And there is an interface capable of interacting with this superposition state, namely the human mind. What we mean is that this superposition state is the concept `The chair in spot A or in spot B'. This is not an object but a concept, and our human mind can conceptually interact with it and
hence act as an interface for it.'' \cite[p. 44]{Aerts10}} \end{quotation}

\noindent Aerts is able to present an original representation of quantum physical reality which goes beyond the measurement problem presenting an hypothesis which might be able to provide an interesting answer to the superposition problem, namely, that superposition states describe the emergence of new meanings. 

Another interpretation which, we believe, attempts to go beyond the mere reference to measurement outcomes is the transactional interpretation proposed by Kastner. In analogous fashion to the conceptuality interpretation, the Possibilist Transactional Interpretation (PTI) discusses the possibility to consider physical reality beyond a strict reference to classical space-time. According to Kastner \cite[p. 147]{Kastner12}: ``PTI is a realist interpretation which, in its strong form, takes the physical referent for quantum states to be ontologically real possibilities existing in a pre-spacetime realm, where the latter is described by Hilbert Space. These possibilities are taken as real because they are physically efficacious, leading indeterministically to transactions which give rise to the empirical events of the spacetime theatre.'' This is also closely related to Aerts proposal where quantum particles are outside space-time, however, Kastner presents us with a different picture of what is really going on.

\begin{quotation}
\noindent {\small ``[...] the dynamical possibilities referred to by state vectors in PTI are Heisenbergian `potentia,' which are less real than events in the actual world, yet more real than mere thoughts or imaginings or conceivable events. [...] Under PTI, the realist use of the term `possible' or `potential' refers to physical possibilities; that is, entities which can directly give rise to specific observable physical phenomena based on a realized transaction. This is distinct from the common usage of the term `possible' or `possibility' to denote a situation or state of affairs which is merely conceivable or consistent with physical law. So, in general, `possibilities' in PTI are entities underlying specific individual events rather than collective, universal sets of events.'' [{\it Op. cit.}, p. 149]} \end{quotation}

\noindent Both Aerts and Kastner's interpretations might be capable of answering the questions we have posed through this article by providing a physical representation which explains what QM is really talking about. A deeper and more exhaustive account of these interpretations exceeds the scope of the present paper, a task we leave for a future work. Let us finish this section by addressing our own account of QM. 

Our approach to QM begins by discussing the extension of the notion of physical reality beyond the limits of the actual realm taking into account the famous definition of an {\it element of physical reality} discussed in the EPR paper \cite{EPR}. According to it:  {\it if, without in any way disturbing a system, we can predict with certainty (i.e., with probability equal to unity) the value of a physical quantity, then there exists an element of reality corresponding to that quantity.} As remarked by Aerts and Sassoli de Bianchi \cite[p. 20]{AertsSassoli17}: ``the notion of `element of reality' is exactly what was meant by Einstein, Podolsky and Rosen, in their famous 1935 article. An element of reality is a state of prediction: a property of an entity that we know is actual, in the sense that, should we decide to observe it (i.e., to test its actuality), the outcome of the observation would be certainly successful.'' Indeed, certainty, taken as the condition of possibility to make reference to the actual realm, has been up to the present the restrictive constraint of what can be considered as part of physical reality. Our redefinition stays close the relation imposed between predictive statements and physical reality, but leaves aside both the actualist constraint imposed by certainty ---restricting existence only to probability equal to unity--- and the strict focus in the process of measurement  ---which should be only regarded as confirming or disconfirming a specific prediction of the theory. Taking into account these general remarks we have proposed in \cite{deRonde16a} the following generalization:\\

\noindent {\it {\bf Generalized Element of Physical Reality:} If we can predict in any way (i.e., both probabilistically or with certainty) the value of a physical quantity, then there exists an element of reality corresponding to that quantity.}\\

\noindent When considering this redefinition the problem is clearly framed: {\it we need to find a set of physical concepts that are capable of being statistically defined in objective terms.} This means, to find a notion that makes reference to a moment of unity that is not defined in terms of {\it yes-no experiments} (as it is the case of objects possessing classical properties), but is defined instead in terms of a {\it probabilistic measure}. We need to develop a conceptual scheme which is consistent with the quantum formalism and is able to explain what QM is talking about beyond the orthodox reference to measurement outcomes and mathematical structures. In turn, this new non-classical scheme must be also capable of explaining in a convincing natural manner the main features brought in by the orthodox quantum formalism: contextuality, interaction of quantum superpositions, entanglement, etc. 

Our proposal begins with the definition of a mode of existence, ontological potentiality, completely independent of the actual realm. It continues by defining two key notions, namely, immanent power and potentia. According to representational realism: {\it physis} is represented in different ways.\footnote{There is in our neo-Spinozist account an implicit ontological pluralism of {\it multiple representations} which can be related to {\it one reality} through a {\it univocity principle}. This is understood in analogous manner to how Spinoza considers in his immanent metaphysics the {\it multiple attributes} as being expressions of the same {\it one single substance}, namely, nature (see \cite{deRonde14, deRonde16b}). Our non-reductionistic answer to the problem of inter-theory relation escapes in this way the requirement present in almost all interpretations of QM which implicitly or explicitly attempt to explain the formalism in substantialist atomistic terms. We believe there might be an interesting connection between our neo-Spinozist approach and the `multiplex realism' recently proposed by Aerts and Sassoli de Bianchi \cite{AertsSassoli15}. Due to the limited space of this paper we leave this particular analysis and comparison for a future work.} So while classical theories represent reality in terms of `particles', `waves' and `fields' all of which exist in the actual realm, QM describes reality in terms of {\it immanent powers} with definite {\it potentia} which exist within a potential realm. This potential realm is not defined ---like the actual realm--- in terms of the Aristotelian PE, PNC and PI. Its definition makes explicit use of the principle of indetermination, the superposition principle and the quantum postulate which might be thought as a principle of difference. Our non-reductionistic approach allows us to concentrate in considering quantum theory independently of the quantum to classical limit. This physical representation of QM can be condensed in the following seven postulates which contain the relation between our proposed new physical concepts and the orthodox formalism of the theory.

\begin{enumerate}

{\bf \item[I.] Hilbert Space:} QM is mathematically represented in a vector Hilbert space.

{\bf \item[II.] Potential State of Affairs (PSA):} A specific vector $\Psi$ with no given mathematical representation (basis) in Hilbert space represents a PSA; i.e., the definite potential existence of a multiplicity of {\it immanent powers}, each one of them with a specific {\it potentia}.

{\bf \item[III.] Quantum Situations, Immanent Powers and Potentia:} Given a PSA, $\Psi$, and the context or basis, we call a quantum situation any superposition of one or more than one power. In general given the basis $B= \{ | \alpha_i \rangle \}$ the quantum situation $QS_{\Psi, B}$ is represented by the following superposition of immanent powers:
\begin{equation}
c_{1} | \alpha_{1} \rangle + c_{2} | \alpha_{2} \rangle + ... + c_{n} | \alpha_{n} \rangle
\end{equation}

\noindent We write the quantum situation of the PSA, $\Psi$, in the context $B$ in terms of the ordered pair given by the elements of the basis and the coordinates in square modulus of the PSA in that basis:
\begin{equation}
QS_{\Psi, B} = (| \alpha_{i} \rangle, |c_{i}|^2)
\end{equation}

\noindent The elements of the basis, $| \alpha_{i} \rangle$, are interpreted in terms of {\it immanent powers}. The coordinates of the elements of the basis in square modulus, $|c_{i}|^2$, are interpreted as the {\it potentia} of the power $| \alpha_{i} \rangle$, respectively. Given the PSA and the context, the quantum situation, $QS_{\Psi, B}$, is univocally determined in terms of a set of powers and their respective potentia. (Notice that in contradistinction with the notion of {\it quantum state} the definition of a {\it quantum situation} is basis dependent and thus intrinsically contextual.)

{\bf \item[IV.] Elementary Process:} In QM we only observe in the actual realm discrete shifts of energy (quantum postulate). These discrete shifts are interpreted in terms of {\it elementary processes} which produce actual effectuations. An elementary process is the path which undertakes a power from the potential realm to its actual effectuation. This path is governed by the {\it immanent cause} which allows the power to remain potentially preexistent within the potential realm independently of its actual effectuation. Each power $| \alpha_{i} \rangle$ is univocally related to an elementary process represented by the projection operator $P_{\alpha_{i}} = | \alpha_{i} \rangle \langle \alpha_{i} |$.

{\bf \item[V.] Actual Effectuation of an Immanent Power (Measurement):} Immanent powers exist in the mode of being of ontological potentiality. An {\it actual effectuation} is the expression of a specific power within actuality. Different actual effectuations expose the different powers of a given $QS$. In order to learn about a specific PSA (constituted by a set of powers and their potentia) we must measure repeatedly the actual effectuations of each power exposed in the laboratory. (Notice that we consider a laboratory as constituted by the set of all possible experimental arrangements that can be related to the same $\Psi$.) An actual effectuation does not change in any way the PSA. 

{\bf \item[VI.] Potentia (Born Rule):} A {\it potentia} quantifies the intensity of an immanent power which exist (in ontological terms) in the potential realm; it also provides a measure of the possibility to express itself (in epistemic terms) in the actual realm. Given a PSA, the potentia is represented via the Born rule. The potentia $p_{i}$ of the immanent power $| \alpha_{i} \rangle$, in the specific PSA, $\Psi$, is given by:
\begin{equation}
p_{(| \alpha_{i} \rangle, \Psi)} = \langle \Psi | P_{\alpha_{i}} | \Psi \rangle = Tr[P_{ \Psi} P_{\alpha_{i}}]
\end{equation}

\noindent In order to learn about a $QS$ we must observe not only its powers (which are exposed in actuality through actual effectuations) but we must also measure the potentia of each respective power. In order to measure the potentia of each power we need to expose the $QS$ statistically through a repeated series of observations. The potentia, given by the Born rule, coincides with the probability frequency of repeated measurements when the number of observations goes to infinity.

{\bf \item[VII.]  Potential Effectuations of Immanent Powers (Schr\"odinger Evolution):} Given a PSA, $\Psi$, powers and potentia evolve deterministically, independently of actual effectuations, producing {\it potential effectuations} according to the following unitary transformation:
\begin{equation}
i \hbar \frac{d}{dt} | \Psi (t) \rangle = H | \Psi (t) \rangle
\end{equation}

\noindent Where $H$ is the Hamiltonian. 
While {\it potential effectuations} evolve according to the Schr\"odinger equation, {\it actual effectuations} are particular expressions of each power (that constitutes the PSA, $\Psi$) in the actual realm. The ratio of such expressions in actuality is determined by the potentia of each power.
\end{enumerate}

\smallskip 

\noindent Let us now continue to analyze in more detail some important aspects of our approach to QM:\smallskip

{\it The potential state of affairs as a set of immanent powers with definite potentia.} Our choice to develop an ontological realm of potentiality absolutely independent of the actual realm of existence implies obviously the need to characterize this realm in an independent manner to classical physical concepts such as `particles', `waves' and `fields' ---notions which are defined in strict relation to the actual mode of existence and the principles that define it. According to our viewpoint, while classical physics talks about systems with definite properties (`particles', `waves' and `fields'), QM talks about the existence of powers with definite potentia. While the classical representation of sets of systems with definite properties can be subsumed under the notion of an {\it actual state of affairs}, QM provides a representation in terms of a {\it potential state of affairs}. This representation seeks on the one hand, to define concepts in a systematic categorical manner avoiding metaphorical discourse, and on the other hand, understand QM ---what it really talks about, the experience it implies--- through these new concepts in an intuitive manner. Several examples have been already discussed in \cite{deRonde16a}. In other words, we need to create a new way of thinking, with new concepts which allow us to define clearly what is observable according to the theory of quanta.  

\smallskip

{\it The existence and interaction of quantum possibilities.} The need to consider quantum possibilities as part of physical reality is supported, in the first place, by the fact that quantum probability resists an ``ignorance interpretation''. The fact that the quantum formalism implies a non-Kolmogorovian probability model which is not interpretable in epistemic terms is a well known fact within the foundational literature since Born's interpretation of the quantum wave function \cite{Redei12}.\footnote{It is true that QBism does provide a subjectivist interpretation of probability following the Bayesian viewpoint, however, this is done so at the price of denying the very need of an interpretation for QM. See for a detailed analysis: \cite{deRonde16a, deRonde16b}. Also the hidden measurement approach by Aerts provides an epistemic interpretation of quantum  probability but in this case, instead of considering the quantum system alone, the approach focuses on the measurement interaction between system and apparatus \cite{AertsSassoli15, AertsSassoli17}.} But more importantly, the quantum mechanical formalism implies that projection operators can be understood as {\it interacting} and {\it evolving} \cite{RFD14}. In classical mechanics the mathematical and conceptual levels are interrelated in such a coherent manner that it makes perfect sense to relate the mathematical description, namely, `a point in phase space evolves according to Newton's equation of motion', with the conceptual representation, namely, `the trajectory of a particle moving in absolute Newtonian space-time'. But in QM, while the interaction and evolution of projection operators is represented quantitively through the mathematical formalism we still lack a conceptual qualitative representation of what projection operators really mean. The {\it interaction} in terms of entanglement, the {\it evolution} in terms of the Schr\"odinger equation of motion and the {\it prediction} of quantum possibilities in statistical terms through the Born rule are maybe the most important features which point towards the need of developing an ontological idea of possibility which is truly independent of actuality. This development is not a mathematical one; rather, it is a metaphysical or conceptual enterprise. 

\smallskip

{\it The intensity of quantum possibilities.} Another important consequence of the ontological perspective towards quantum possibilities relates to the need of reconsidering the binary existencial characterization of properties in terms of an homomorphic relation to the binary Boolean elements $\{ 0,1 \}$ ---a relation which amounts to the valuation of propositions in truth tables. Our proposed extension of the notion of {\it element of physical reality} escapes the actualist characterization of existence in terms of certitude (probability = 1) and considers ``right from the start'' the quantum probabilistic measure in objective terms. This move implies the development of existence beyond the gates of certitude and the complementary need of characterizing the basic elements of our ontology ---namely, immanent powers--- in intensive terms; i.e. as relating to a value which pertains to the interval $[0,1]$. In this way, each {\it immanent power} has an intensive characterization which we call {\it potentia}. We could say that, unlike properties that pertain to systems either exist or do not exist (i.e., they are related either to 1 or 0), immanent powers have a more complex characterization which requires, apart from its binary relation to existence, a number pertaining to the closed interval $[0,1]$ which specifies its (potential) existence in an intensive manner. It is through the introduction of an intensive mode of existence that we can understand quantum probability as describing an objective feature of QM ---rather than epistemic ignorance about an actual state of affairs.

\smallskip 

{\it Immanent powers and contextuality.} It is important to notice that the intensive characterization of immanent powers allows us to escape Kochen-Specker contextuality \cite{deRonde17b} and restore a global valuation to all projection operators of a quantum state, $\Psi$. By removing the actualist binary reference of classical properties to $\{ 0,1 \}$, and implementing instead an intensive valuation of projection operators to an element within the interval $[0,1]$ we are able, not only to bypass Kochen-Specker theorem \cite{KS}, but also to restore ---through a {\it global intensive valuation}--- an objective representation of the elements the theory talks about. Powers are non-contextual existents which can be defined univocally and globally for any given quantum state $\Psi$ (i.e., a particular potential state of affairs). In this way, just like in the case of classical physics, quantum contextuality can be understood as exposing the epistemic incompatibility of measurement situations and outcomes (see for a detailed discussion and analysis: \cite{deRonde16a, deRonde17b, RFD14}).

\smallskip 

{\it The contradiction of quantum possibilities.} Some quantum superpositions of the ``Schr\"odinger cat type'' \cite{Schr35} constituted by two contradictory terms, e.g. `$| + \rangle$' and `$| - \rangle$', present a difficult problem for those who attempt to describe the theory in terms of particles with definite non-contradictory properties. Indeed, as discussed in \cite{daCostadeRonde13}, while the first term might relate to the statement `the atom possesses the property of being decayed' the second term might relate to the statement `the atom possesses the property of not being decayed'.  Obviously, an atom cannot be `decayed' and `not decayed' at the same time ---just like a cat cannot be `dead' and `alive' simultaneously. Any classical physical object ---an atom, a cat, a table or a chair---, by definition, can only posses non-contradictory properties. Physical objects have been always ---implicitly or explicitly--- defined since Aristotle's metaphysics and logic in terms of the principles of existence, non-contradiction and identity. However, regardless of the manner in which objects are defined in classical physics, QM allows us to predict through the mathematical formalism how these terms will interact and evolve in different situations. The realist attitude is of course to consider that the formalism, and in particular quantum superpositions, are telling us something very specific about physical reality. Something which in fact has allowed us to enter the new technological era of quantum information processing. This is also why one might consider Schr\"odinger's analysis as an {\it ad absurdum} proof of the impossibility to describe quantum superpositions in terms of classical notions (i.e., particles, waves or fields). Our representation in terms of powers with definite potentia attempts to fill this conceptual lacuna. 

\smallskip  

{\it The relation and independence of immanent powers with respect to the actual realm.} Immanent powers have an independent potential existence with respect to the actual realm. Measurement outcomes are not what potential powers attempt to describe. It is exactly the other way around ---at least for a representational realist. For the realist, measurement outcomes are only expressions of a deeper moment of unity which requires a categorical definition. This is completely analogous to the classical case in which the observation of a table at three-fourteen (seen in profile) or the table at three-fifteen (seen from the front) are both particular expressions which find their moment of unity in the notion of `table'. However, we still need to provide an answer to the measurement problem and explain in which manner quantum superpositions (in formal terms), and powers with definite potentia (in conceptual terms), relate to actual effectuations. In \cite{deRonde17}, we have provided a detailed analysis of our understanding of how the measurement process should be understood in QM. Within our approach, the quantum measurement process is modeled in terms of the Spinozist notion of {\it immanent causality}. The immanent cause allows for the expression of effects remaining both in the effects and its cause. It does not only remain in itself in order to produce, but also, that which it produces stays within. Thus, in its production of actual effects the potential does not deteriorate by becoming actual ---as in the case of the hylomorphic scheme of causal powers (see section 1, p. 4 of \cite{deRonde17}).\footnote{For a more detailed discussion of the notion of immanent cause we refer to \cite[Chapter 2]{Melamed}.} Immanent powers produce, apart from actual effectuations, also {\it potential effectuations} which take place within potentiality and remain independent of what happens in the actual realm. Within our model of measurement, while potential effectuations describe the ontological interactions between immanent powers and their potentia ---something known today as {\it entanglement}---, actual effectuations are only epistemic expressions of the potentia of powers. Actualities are only partial expressions of powers. Just like when observing a dog, a table or a chair we only see a partial perspective of the object but never the object itself, measurement outcomes expose only a partial account of the potentia of powers. \smallskip  

``Getting rid of the ghost of Schr\"odinger's cat'' is a phrase used by Griffiths \cite{Griffiths13} which we find very appropriate to describe a common animosity present in the foundational literature against quantum superpositions. This general uncomfortable feeling with respect to superpositions can be witnessed in the negative adjectives which accompany their definition in many articles:  ``embarrassing'', ``weird'', ``strange'', ``spooky'', etc. The truth is that Schr\"odinger's creation is indeed a strange kind of ``zombie-cat'' dead and alive at the same time; a creature difficult to picture or imagine ---closer to the mysterious Cheshire cat from {\it Alice in Wonderland} than to a particle. And this might be the reason why most interpretations have attempted, either to get rid of what they consider to be a ``ghost'' or simply argue that in ``classical reality'' ghosts do not exist. Instead of trying to fight or destroy quantum superpositions, we believe it would be much more interesting for the philosophy of physics community to try to understand them. The just mentioned accounts of QM show that it is possible to discuss beyond the actualist classical representation of physics and provide a non-empiricist explanation of what the theory of quanta is really talking about. Even though these are just starting points within lines of research which require a lot more work to be done, we believe that it would be worth while to take these approaches as standpoints in order to discuss the questions that have been posed in the present article regarding quantum superpositions.

\section{Conclusion}

In the first part of this paper we have argued that, according to our representational realist stance, mathematical structures are not capable of providing by themselves a physical representation of a theory. Physical theories are also necessarily related to a network of specific physical concepts ---defined in a systematic manner. Mathematical structures provide a quantitative understanding about phenomena. However, this mathematical account does not provide the qualitative understanding produced by physical notions. Thus, until we do not find a conceptual scheme which coherently relates to the orthodox formalism  and expresses the moment of unity of the multiple phenomena we will no be able to say we have understood QM. We have also argued that ---in principle--- it should be possible to come up with a physical network of concepts that take into account the non-classical features of QM. The price to pay maybe the abandonment of the attempt to explain the theory of quanta in terms of the ``common sense'' classical Newtonian actualist metaphysics. This abandonment might be the condition of possibility that could allow us to construct a new non-classical metaphysical scheme with physical concepts specifically designed in order to account for the orthodox formalism of QM. We know of no reason to believe this is not doable.

In the second part of the paper we have criticized the dogmatic constrains of the orthodox literature in order to discuss and analyze the quantum formalism in general, and in particular, quantum superpositions. From a representational realist stance, we have argued in favor of the necessity to consider a conceptual representational level defined through a metaphysical architectonic which describes quantum superpositions beyond the reference to mathematical structures and measurement outcomes. We have also provided a formal redefinition of quantum superpositions which takes into account their contextual nature. We presented three necessary conditions for any objective physical representation and introduced the superposition problem which, contrary to the measurement problem, focuses on the conceptual interpretation of superpositions themselves. Furthermore, we provided several arguments which point in the direction of considering quantum superpositions as real physical existents. In the final part of the paper we discussed, firstly, why the actualist interpretations fail to provide such a representation, and secondly, the possibilities implied by certain interpretations such as those of Aerts and Kastner in order to produce a physical representation beyond the constraints imposed by the classical actualist representation of physical reality. We ended the article considering our own approach to quantum mechanics which takes as a standpoint the generalization of the notion of element of physical reality.

\section*{Acknowledgements} 

I wish to thank two anonymous reviewers for their careful reading of my manuscript and their many insightful comments and suggestions. This work was partially supported by the following grants: FWO project G.0405.08 and FWO-research community W0.030.06. CONICET RES. 4541-12 and the Project PIO-CONICET-UNAJ (15520150100008CO) ``Quantum Superpositions in Quantum Information Processing''.


\end{document}